\documentclass[twocolumn]{aastex701} 
\usepackage{graphicx,amsmath,amssymb}
\usepackage{natbib}
\usepackage{hyperref}
\usepackage{mathrsfs}
\usepackage{xcolor}
\usepackage{bm}
\usepackage{ulem} 

\newcommand{\be}{\begin{equation}}
\newcommand{\ee}{\end{equation}}
\newcommand{\ba}{\begin{eqnarray}}
\newcommand{\ea}{\end{eqnarray}}

\newcommand{\MSD}{\langle \Delta X^2(t) \rangle}

\newcommand{\Drlx}{D_{J, {\rm rlx}}}
\newcommand{\DRR}{D_{J, {\rm RR}}}
\newcommand{\Grlx}{\Gamma_{\rm rlx}}
\newcommand{\GRR}{\Gamma_{\rm RR}}
\newcommand{\LFP}{\mathcal{L}_{\text{FP}}}
\newcommand{\LFrac}{\mathbb{D}_J^q}
\newcommand{\JSB}{J_{\text{SB}}}
\newcommand{\Tdiff}{T_{\rm diff}}
\newcommand{\msun}{M_{\odot}}

\def\mnras{Mon. Not. R. Astron. Soc.}
\def\apj{Astrophys. J.}
\def\apjl{Astrophys. J. Lett.}

\def\aap{Astron. Astrophys.}

\shorttitle{Fractional Dynamics and Non-Local Transport in Galactic Nuclei}
\shortauthors{Amaro Seoane}

\begin{document}

\title{Fractional Dynamics in Galactic Nuclei\\ Non-Local Transport, Transient Phenomena and the Nullification of the Schwarzschild Barrier}

\author[orcid=0000-0003-3993-3249,gname='Pau',sname='Amaro Seoane']{Pau Amaro Seoane}
\affiliation{Universitat Politècnica de València, C/Vera s/n, València, 46022, Spain}
\affiliation{Max Planck Institute for Extraterrestrial Physics, Giessenbachstra{\ss}e 1, Garching, 85748, Germany}
\email[show]{amaro@upv.es}

\begin{abstract}
We investigate the application of fractional calculus to model stellar dynamics, focusing on Resonant Relaxation (RR) near a supermassive black hole (SMBH). Standard theories use the local Fokker-Planck (FP) equation, restricted to Gaussian processes under the Central Limit Theorem (CLT). We argue this is inadequate for RR. We demonstrate that gravitational interactions inherently produce infinite variance in stochastic torques, violating the CLT. Consequently, RR is governed by the Generalized Central Limit Theorem (GCLT) and constitutes a superdiffusive L\'{e}vy flight. We apply the space-fractional Fokker-Planck equation (FFPE), utilizing non-local operators, to explore resolutions to observational discrepancies. In transient regimes, the FFPE predicts immediate, linear flux ($\Gamma(t) \propto t$), consistent with high Tidal Disruption Event (TDE) rates in post-starburst galaxies, whereas local FP models predict significant exponential delay. Furthermore, we demonstrate analytically that non-local integral operators permit ``barrier jumping,'' bypassing bottlenecks like the Schwarzschild Barrier (SB), which local models interpret as severely suppressing Extreme Mass-Ratio Inspiral (EMRI) rates. We present proof-of-concept $N$-body simulations that confirm non-local RR transport, although the resolution must be improved to rule out enhanced Two-Body Relaxation in the small-N setup. The fractional framework offers a compelling alternative description for non-local transport, potentially resolving TDE and EMRI rate questions.
\end{abstract}


\section{Introduction}

The dynamics of stars and stellar-mass black holes (sBHs) in the vicinity of a supermassive black hole (SMBH) govern the rates of astrophysically significant events \citep[see e.g.][and references therein]{Amaro-SeoaneLRR}. These include Extreme Mass-Ratio Inspirals (EMRIs), which are primary sources for space-borne gravitational wave detectors, and Tidal Disruption Events (TDEs). Accurately modeling the transport processes in these dense stellar environments is therefore fundamental to predicting detection rates and interpreting observations.

In the near-Keplerian potential of a galactic nucleus, the dominant mechanism for angular momentum transport is Resonant Relaxation (RR), first described by \citet{RT96}. RR arises from the coherent torques exerted by the fluctuating stellar background potential on individual orbits. These torques persist over the orbital precession timescale, leading to large, correlated changes in angular momentum.

{The theoretical understanding of RR, both scalar (changes in magnitude of angular momentum) and vector (changes in orientation, VRR), has advanced significantly through the application of kinetic theory and secular approximations. These approaches often utilize the Balescu-Lenard formalism or related kinetic equations to describe the long-term evolution driven by resonant encounters and correlated noise from potential fluctuations \citep{FouvryBarOr2018, BarOrFouvry2018, FouvryEtAl2019_b}. This framework has enabled the calculation of diffusion coefficients from first principles \citep{BarOrFouvry2018} and has been applied to complex phenomena such as the coupling of VRR with Lidov-Kozai dynamics \citep{HamersEtAl2018_b}. Furthermore, sophisticated numerical methods based on secular dynamics and multipole expansions have been developed to study these processes efficiently \citep{BarOrEtAl2021}. While much of this work focuses on systems amenable to standard kinetic treatments, specific regimes, such as 1D homogeneous systems or certain frequency profiles, exhibit phenomena like kinetic blocking where standard relaxation is suppressed \citep{FouvryEtAl2019, FouvryEtAl2019_c}.}

The statistical nature of RR deviates fundamentally from classical two-body relaxation (subscript ``rlx'' from here on), which is typically characterized by Brownian motion. {We propose that RR exhibits characteristics consistent with a L\'{e}vy flight. This type of random walk is characterized by trajectories featuring many short steps interspersed with occasional, large jumps, resulting from a step-length probability distribution with a heavy, power-law tail. We provide a formal theoretical justification for this behavior in Section~\ref{sec:justification}.}

{If the distribution of angular momentum step sizes indeed exhibits a heavy power-law tail (see Eq.~\ref{eq:levy_tail}),} This behavior signifies a superdiffusive process, where the mean squared displacement (MSD) grows faster than linearly with time,
\begin{equation} \label{eq:msd_anomalous}
\MSD = K_\alpha t^\alpha,
\end{equation}
\noindent
with $1 < \alpha < 2$. The divergence of the second moment of the step size distribution implies that transport is inherently non-local; large jumps dominate the dynamics.

Despite the recognition of RR as a highly efficient transport mechanism, the standard approach in stellar dynamics often models its evolution using the classical Fokker-Planck (FP) equation (Eq.~\ref{eq:FP_local}). This formalism employs local differential operators and is mathematically restricted to processes governed by the Central Limit Theorem (CLT), yielding Gaussian propagators.

Significant advancements were made in understanding how relativistic precession affects RR. Direct-summation $N-$body simulations heuristically suggested a suppression of RR diffusion at low angular momentum, dubbed the ``Schwarzschild Barrier'' \citep[SB,][]{MerrittEtAl11}. {This effect was also observed in subsequent numerical studies \citep[e.g.,][]{HamersEtAl2014}.} This phenomenon was subsequently addressed by \citet{BarOrAlexander2014} in terms of Adiabatic Invariance (AI): As put forward by the authors, when the precession period is shorter than the RR coherence time, the coherent torques are averaged out. They developed the $\boldsymbol{\eta}$-formalism to calculate effective diffusion coefficients (DCs) by modeling the background potential as correlated noise, leading to the conclusion that the SB severely inhibits EMRI rates \citep{BarOrAlexander2016}.

Whilst these approaches provided foundational insights into the interplay between precession and RR, we note that they rely on the assumption that the background potential fluctuations are Gaussian. As explicitly stated in Section 2.2 of \citet{BarOrAlexander2014}, they invoke the CLT to describe the superposition of forces from $N\gg1$ stars as time-dependent Gaussian random variables. {This approach is standard, provided the variance of the individual stochastic contributions is finite. However, as argued in Section~\ref{sec:justification}, the variance of gravitational torques may be infinite. In such cases, the standard CLT does not apply, even for $N\gg1$.} This assumption inherently restricts the analysis to systems with finite variance in the stochastic steps, resulting in a local FP description.

{Furthermore, subsequent theoretical work based on kinetic theory, such as \citet{BarOrFouvry2018}, also adopted local formalisms. These models successfully reproduced the eccentricity diffusion coefficients measured in some numerical simulations, such as those analyzed in \citet{HamersEtAl2014}. This agreement between local theory and simulations warrants discussion. If RR is indeed a L\'{e}vy flight, the agreement might indicate that the local models capture the average behavior well in the specific regimes studied, but may fail in scenarios sensitive to the tails of the distribution, such as transport across barriers or transient phenomena.}

We analyze the limitations of applying a local formalism to a process characterized by L\'{e}vy flights. A L\'{e}vy flight is a stochastic process defined by a step-length distribution $P(\Delta J)$ with a power-law tail $P(\Delta J) \sim |\Delta J|^{-1-q}$ (Eq.~\ref{eq:levy_tail}). A mathematical consequence is the divergence of the second moment (variance), $\langle (\Delta J)^2 \rangle \to \infty$ \citep[see Sec 3.5 of][]{MetzlerKlafter2000}. Such processes are governed by the L\'{e}vy-Gnedenko Generalized Central Limit Theorem (GCLT), which states that the sum of these steps converges to a L\'{e}vy stable distribution, not a Gaussian \citep[see Sec 2.1 and 3.7.2 of][]{MetzlerKlafter2000}. Applying a local formalism, which is mathematically predicated on the standard CLT and the existence of a finite variance, to this GCLT process presents a fundamental inconsistency. The local model artificially imposes a Gaussian propagator (Eq.~\ref{eq:propagator_gaussian}) with exponentially decaying tails onto a dynamic characterized by large, non-local jumps.

This suppression of high-probability large jumps leads to significant discrepancies in predicting event rates. Local models predict a bottleneck at the SB, suggesting transport across the barrier relies solely on slower two-body relaxation. {This interpretation is based on the observation of the SB in numerical simulations \citep{MerrittEtAl11, HamersEtAl2014}. We revisit the interpretation of these findings in light of the non-local framework in Section~\ref{sec:schwarzschild}.} Furthermore, in transient regimes, such as the refilling of the loss cone \citep{FR76} in post-starburst (E+A) galaxies, observations indicate high TDE rates \citep{ArcaviEtAl2014,FrenchEtAl2016,LawSmithEtAl2017}, implying rapid replenishment. Local models predict a significant delay, as their inherent step-by-step diffusive nature cannot reproduce this rapid response.

In this work, we introduce a framework based on fractional calculus \citep[see e.g.][]{Meerschaert2011StochasticMF} to model RR as a L\'{e}vy flight. Superdiffusion requires the use of the space-Fractional Fokker-Planck Equation \citep[space-FFPE, see e.g.][]{BarkaiEtAl2000}. This formalism employs integro-differential operators (e.g., Riemann-Liouville derivatives) that correctly capture the non-local transport and yield the appropriate L\'{e}vy stable propagators.

We apply this framework to the aforementioned astrophysical problems. We demonstrate that the non-local nature of the fractional operators {can potentially resolve the discrepancies inherent in local models}. In the transient regime, the fractional framework predicts an immediate, linear increase in the TDE flux. Furthermore, we {show analytically} that the non-local dynamics allow stars to bypass the region of suppressed diffusion via a ``barrier jumping'' mechanism. We corroborate this theoretical finding with direct-summation $N$-body simulations incorporating post-Newtonian corrections. These simulations {suggest} efficient transport across the SB. This combined theoretical and numerical evidence {suggests} that the SB does not constitute an actual barrier to RR-driven transport, {potentially alleviating} a major theoretical bottleneck for EMRI production.

\section{{L\'{e}vy Flights and RR}}
\label{sec:justification}

{The application of a fractional diffusion framework to Resonant Relaxation (RR) is rooted in the statistical mechanics of the underlying gravitational interactions. In this section we elaborate this argument and demonstrate that the standard assumptions underpinning the local Fokker-Planck formalism are violated by the $1/r^2$ nature of gravity, necessitating a description based on L\'{e}vy flights and the Generalized Central Limit Theorem (GCLT).}

{\subsection{The Breakdown of the Central Limit Theorem in Gravity}}

{The standard Fokker-Planck equation describes Markov processes where the stochastic fluctuations are Gaussian. This description relies on the Central Limit Theorem (CLT), which states that the sum of many independent and identically distributed (IID) random variables converges to a Gaussian distribution, provided the variance ($\sigma^2$) of the individual variables is finite.}

{However, the gravitational field in a stellar system exhibits fundamentally different statistics. The foundational analysis by \citet{Holtsmark1919} and \citet{Chandrasekhar1943} established that the probability distribution of the gravitational force $\vec{F}$ is not Gaussian. This is formally derived by analyzing the characteristic function (the Fourier transform of the PDF), $\Phi(\vec{k}) = \langle e^{i \vec{k} \cdot \vec{F}} \rangle$. For a 3D isotropic distribution of stars (density $n$), the characteristic function is:}
\begin{equation} \label{eq:char_func_holtsmark}
\Phi(\vec{k}) = \exp(-C n |\vec{k}|^{3/2}),
\end{equation}
\noindent
{where $C$ is a constant. This is the Holtsmark distribution, a specific type of L\'{e}vy $\alpha$-stable distribution with stability index $\alpha=3/2$. The non-analytic dependence on $|\vec{k}|$ (specifically, the exponent $3/2 < 2$) is a direct consequence of the $1/r^2$ force law, which allows the total force to be dominated by rare, strong interactions with nearby stars. The resulting PDF possesses a heavy power-law tail:}
\begin{equation} \label{eq:holtsmark_tail}
P(F) \sim F^{-5/2}, \quad (F \to \infty).
\end{equation}

{\subsection{Infinite Variance of Gravitational Torques}}

{We now demonstrate that the torques governing RR inherit this heavy-tailed behavior, leading to a divergence of the variance. We analyze the distribution of torques $P(\tau)$ using the nearest-neighbor approximation, which dominates the tail of the distribution.}

{In a 3D system, the probability density for the nearest neighbor distance $R$ scales, for small $R$, as $P(R) \propto R^2$. The torque magnitude scales as $\tau \propto 1/R^2$. We perform a change of variables from $R$ to $\tau$. Let $\tau = C'/R^2$. Then $R \propto \tau^{-1/2}$, and the Jacobian is $|dR/d\tau| \propto \tau^{-3/2}$. The PDF of the torque is $P(\tau) = P(R(\tau)) |dR/d\tau|$:}
\begin{align}
P(\tau) \propto R^2 \, \tau^{-3/2} \propto (\tau^{-1/2})^2\, \tau^{-3/2} = \tau^{-5/2}.
\end{align}
\noindent
{The distribution of gravitational torques exhibits a heavy tail $P(\tau) \sim \tau^{-(1+\alpha)}$ with index $\alpha=3/2$. We now evaluate the second moment (variance):}
\begin{equation}
\langle \tau^2 \rangle = \int_0^{\infty} \tau^2 P(\tau) d\tau \propto \int^{\infty} \tau^2 \, \tau^{-5/2} d\tau = \int^{\infty} \tau^{-1/2} d\tau.
\end{equation}
\noindent
{This integral diverges. The variance of the gravitational torque is infinite. More generally, for a L\'{e}vy stable distribution with index $\alpha$, all fractional moments $\langle |\tau|^\theta \rangle$ diverge if $\theta \ge \alpha$.}

{\subsection{The Generalized Central Limit Theorem and Fractional Kinetics}}

{The divergence of the variance fundamentally invalidates the standard CLT. Consequently, the assumption that the superposition of torques results in a Gaussian process (as invoked in, e.g., \citealt{BarOrAlexander2014} based on $N\gg1$) is mathematically inconsistent with the statistics of the gravitational field.}

{When the variance is infinite, the appropriate statistical framework is the Generalized Central Limit Theorem (GCLT) \citep{MetzlerKlafter2000}.}

\noindent
{GCLT Statement:} {The sum of IID random variables with a power-law tail $P(X) \sim |X|^{-(1+\alpha)}$, where $0 < \alpha < 2$, converges to a L\'{e}vy $\alpha$-stable distribution.}

{In RR, the angular momentum changes $\Delta J$ accumulate over time. Assuming the coherence time is not strongly anti-correlated with the torque magnitude, the distribution of $\Delta J$ inherits the heavy tail $P(\Delta J) \sim |\Delta J|^{-5/2}$. Since $\alpha=3/2 < 2$, the GCLT applies. The resulting stochastic process is a L\'{e}vy flight, characterized by superdiffusion and scale invariance.}

{The kinetic equation governing the evolution of the PDF $f(J,t)$ for a L\'{e}vy flight is fundamentally different from the standard Fokker-Planck equation (which corresponds to $\alpha=2$). A Markovian process converging to a L\'{e}vy stable distribution must be governed by the space-Fractional Fokker-Planck Equation (FFPE):}
\begin{equation}
\frac{\partial f(J,t)}{\partial t} = \gamma \frac{\partial^\alpha f(J,t)}{\partial |J|^\alpha}.
\end{equation}
\noindent
{The spatial operator is the Riesz fractional derivative, which is a non-local integro-differential operator. Its non-locality is essential to capture the long-range jumps of the L\'{e}vy flight. The connection to the underlying statistics is evident in the Fourier domain. The Fourier transform of the Riesz derivative is:}
\begin{equation}
\mathcal{F}_k \left[ \frac{\partial^\alpha f(J,t)}{\partial |J|^\alpha} \right] = -|k|^\alpha \mathcal{F}_k[f(J,t)].
\end{equation}
\noindent
{The symbol $-|k|^\alpha$ is the generator of the L\'{e}vy stable process. This structure, specifically with $\alpha=3/2$, directly corresponds to the characteristic function derived from the gravitational interaction (Eq.~\ref{eq:char_func_holtsmark}). }

\section{Anomalous Diffusion and the Fractional Formalism}
\label{sec:formulation}

We establish now the framework required to model anomalous diffusion (see e.g. \citealt{MetzlerKlafter2000}, and \citealt{MetzlerEtAl2014} in the context of Brownian diffusion with Gaussian propagators) in galactic nuclei, derived from the underlying Continuous Time Random Walk (CTRW) model.

We analyze the regimes of anomalous diffusion based on the exponent $\alpha$ in Eq.~(\ref{eq:msd_anomalous}). Subdiffusion, corresponding to $\alpha < 1$, arises from processes with a heavy-tailed waiting time distribution $\psi(t) \sim t^{-1-\beta}$ (where $0 < \beta < 1$). This leads to a diverging mean waiting time, effectively "trapping" particles. This regime is non-Markovian, exhibiting memory effects as discussed in Section~2.3, and could represent stars temporarily caught in resonances. Normal diffusion, $\alpha = 1$, represents the classical limit where the mean squared displacement grows linearly with time. This process has finite variance in both jump lengths and waiting times and is governed by the standard Central Limit Theorem. In our context, this corresponds to classical two-body relaxation over long timescales. Finally, superdiffusion, $1 < \alpha < 2$, arises when the jump length distribution $\lambda(X)$ is heavy-tailed (Eq.~\ref{eq:levy_tail}). This process, known as a L\'{e}vy flight, is characterized by an infinite variance in jump lengths. The MSD grows faster than linearly with time. {Based on the arguments presented in Section~\ref{sec:justification}, we identify this superdiffusive regime as the appropriate description for Resonant Relaxation (RR) in angular momentum space.}

{The coherent torques in RR, persisting for the precession timescale $T_{\text{precess}}$, generate a L\'{e}vy flight characterized by} a probability density function (PDF) $P(\Delta J)$ with a power-law tail,
\begin{equation} \label{eq:levy_tail}
P(\Delta J) \sim |\Delta J|^{-1-q}, \quad 1 < q < 2.
\end{equation}
\noindent
{(Where $q$ corresponds to the L\'{e}vy index $\alpha$ discussed in Section~\ref{sec:justification}, theoretically $q=3/2$.)}

It is worth noting that even two body relaxation exhibits anomalous characteristics in energy space. As detailed by \citet{BarOrKupiAlexander2013}, the energy transition probability $K(\Delta E) \sim |\Delta E|^{-3}$ leads to diverging higher-order moments in the Kramers-Moyal expansion, meaning the standard FP approximation fails on short timescales and energy relaxation proceeds as anomalous diffusion. While this highlights the general limitations of the FP approach, the focus here is on the distinct superdiffusive mechanism of RR in angular momentum space.

\subsection{The CTRW Foundation and the Generalized Master Equation}

The Fractional Fokker-Planck Equations (FFPEs) arise as the hydrodynamic limit of the CTRW model, defined by a jump length distribution $\lambda(X)$ and a waiting time distribution $\psi(t)$. Anomalous diffusion emerges when either $\lambda(X)$ or $\psi(t)$ (or both) are heavy-tailed, such that their characteristic moments diverge \citep[see Sec 3 of][]{MetzlerKlafter2000}. This divergence is the fundamental reason why the standard Fokker-Planck formalism, which relies on the finiteness of the first two moments (as required by the Pawula theorem to truncate the Kramers-Moyal expansion), is insufficient for RR.

The CTRW model generates two distinct classes of anomalous diffusion. First, \textit{subdiffusion} ($\alpha < 1$), which arises from a heavy-tailed $\psi(t)$ (diverging mean waiting time), leads to a \textit{time-fractional} FFPE. This equation, $\partial W / \partial t = {}_{0}D_t^{1-\alpha} \LFP W$, involves a non-local \textit{temporal} operator (the Riemann-Liouville fractional derivative) but retains the standard \textit{local} spatial operator $\LFP$ \citep[see Sec 3.4 and 5.2 of][]{MetzlerKlafter2000}. This models non-Markovian trapping. Second, \textit{superdiffusion} ($1 < \alpha < 2$), which arises from a heavy-tailed $\lambda(X)$ (diverging jump variance), leads to a \textit{space-fractional} FFPE. This equation, $\partial W / \partial t = K^\mu \nabla^\mu W$, involves a local \textit{temporal} operator (is Markovian) but utilizes a non-local \textit{spatial} operator \citep[see Sec 3.5, Eq. 58 of][]{MetzlerKlafter2000}. As we argue RR is a L\'{e}vy flight, this second case is the correct physical description. This distinction is critical, as a subdiffusive model, despite being non-Markovian, would still treat the SB as a local spatial barrier.

This divergence is the fundamental reason why the standard Fokker-Planck formalism, which relies on the finiteness of the first two moments (as required by the Pawula theorem to truncate the Kramers-Moyal expansion), is insufficient for RR.

We can derive the governing FFPE by analyzing the CTRW limit in the Fourier-Laplace (FL) domain \citep{Meerschaert2011StochasticMF}. Let the probability density of the CTRW limit process be $q(x,t)$. This is the probability of being at position $x$ at time $t$, and it results from a sum over all possible jump counts. This can be expressed as an integral over the operational time $u$, $q(x,t) = \int_0^\infty p(x,u) h(u,t) du$, where $p(x,u)$ is the PDF for the $\alpha$-stable L\'{e}vy motion (the jumps) and $h(u,t)$ is the PDF for the inverse $\beta$-stable process (the waiting times) \citep{Meerschaert2011StochasticMF}.

The FL transform (Fourier $x \to k$, Laplace $t \to s$) of the PDF $q(x,t)$ is $\overline{q}(k,s)$. This is the product of the Fourier transform of $p(x,u)$, $\hat{p}(k,u) = \exp(u \psi_A(k))$, and the Laplace transform of $h(u,t)$, $\tilde{h}(u,s) = s^{\beta-1} \exp(-u s^\beta)$, where $\psi_A(k) = (ik)^\alpha$ is the Fourier symbol for the spatial jumps and $\psi_D(s) = s^\beta$ is the Laplace symbol for the waiting times \citep{Meerschaert2011StochasticMF}. Integrating this product yields the FL transform of the CTRW limit PDF \citep{Meerschaert2011StochasticMF},

\begin{equation}
\overline{q}(k,s) = \int_0^\infty e^{u \psi_A(k)} s^{\beta-1} e^{-u \psi_D(s)} du = \frac{s^{\beta-1}}{s^\beta - (ik)^\alpha}.
\end{equation}

\noindent
Rearranging this algebraic relation gives

\begin{equation} \label{eq:FFPE_fourier_laplace}
s^\beta \overline{q}(k,s) - s^{\beta-1} = (ik)^\alpha \overline{q}(k,s).
\end{equation}

Inverting this equation back to the real-space domain term by term yields the FFPE. The left-hand side, $s^\beta \overline{q}(k,s) - s^{\beta-1}$, inverts to the Caputo time-fractional derivative $\partial_t^\beta q(x,t)$, which codes for the power-law waiting times \citep{Meerschaert2011StochasticMF}. The right-hand side, $(ik)^\alpha \overline{q}(k,s)$, inverts to the Riesz-Weyl space-fractional derivative $\mathbb{D}_x^\alpha q(x,t)$, which codes for the power-law L\'{e}vy jumps \citep{Meerschaert2011StochasticMF}. This results in the general FFPE, $\partial_t^\beta q = \mathbb{D}_x^\alpha q$. In the context of Resonant Relaxation, we model a Markovian process (non-heavy-tailed waiting times, so $\beta=1$) with L\'{e}vy flight jumps in angular momentum $J$ (requiring $1 < \alpha < 2$). Setting $\beta=1$ and identifying the spatial variable $x$ with $J$, the distribution function $q(x,t)$ with $f(J,t)$, and the index $\alpha$ with $q$, recovers the space-fractional FFPE utilized in Eq.~(\ref{eq:space_ffpe}).

\subsection{Case 1: Subdiffusion (Non-Local in Time)}
Subdiffusion ($\alpha < 1$) occurs when the waiting time distribution is heavy-tailed, $\psi(t) \sim t^{-1-\beta}$ ($0 < \beta < 1$), leading to diverging mean waiting time (trapping). The jump lengths have finite variance.

In the hydrodynamic limit, $\Psi(k, u) \sim -u^{1-\beta} K_\beta k^2$. Transforming back to the real-time domain yields the Time-FFPE,
\begin{equation} \label{eq:time_ffpe}
\frac{\partial^\beta f(X, t)}{\partial t^\beta} = K_\beta(X) \mathcal{L}_{\text{FP}} f(X, t).
\end{equation}
\noindent
The spatial operator $\mathcal{L}_{\text{FP}}$ is the standard, local Fokker-Planck operator. The temporal operator is the Caputo fractional derivative,
\begin{equation} \label{eq:caputo}
\frac{\partial^\beta f}{\partial t^\beta} = \frac{1}{\Gamma(1-\beta)} \int_0^t \frac{\partial f(X, \tau)}{\partial \tau} (t-\tau)^{-\beta} d\tau.
\end{equation}

\noindent
This operator introduces a memory kernel $(t-\tau)^{-\beta}$, which explicitly represents the non-Markovian nature of the process. In the context of stellar dynamics, this implies that the evolution of the stellar distribution function $f(X, t)$ does not depend solely on its present state. A standard, Markovian process, described by a first-order time derivative $\partial f / \partial t$, assumes the system's future depends only on $f(X, t)$. In contrast, the Caputo derivative (Eq.~\ref{eq:caputo}) computes the current fractional rate of change by integrating the standard rate of change $\partial f(X, \tau)/\partial \tau$ over the system's entire past history (from $\tau=0$ to $\tau=t$). This formalism models subdiffusive processes which arise from trapping effects. In a dense stellar system, this can represent stars temporarily captured in resonances or binary systems, leading to a heavy-tailed waiting time distribution $\psi(t) \sim t^{-1-\beta}$ for diffusive jumps. The kernel contains the persistence of these trapped states; the system's evolution at time $t$ is thus dependent on the full history of $f(X, \tau)$, which defines its non-Markovian character.

\subsection{Case 2: Superdiffusion (Non-Local in Space)}
\label{subsec:superdiffusion_formalism}
Superdiffusion ($1 < \alpha < 2$) occurs when the jump length distribution is heavy-tailed, $\lambda(X) \sim |X|^{-1-q}$ ($1 < q < 2$), leading to diverging variance (L\'{e}vy flights). The waiting times are finite. This scenario corresponds to the Generalized CLT, resulting in L\'{e}vy stable distributions rather than Gaussian distributions.

In the hydrodynamic limit, $\Psi(k, u) \sim -D_q |k|^q$. Transforming back to the real-space domain yields the Space-FFPE. For resonant relaxation in a heterogeneous cusp, we employ the Variable-Order form (VO-FFPE), recognizing that the physical parameters may vary across the cusp,
\begin{equation} \label{eq:space_ffpe}
\frac{\partial f(J, t)}{\partial t} = \mathcal{L}_{\text{Drift}} f + D(J) \mathbb{R}^{q(J)} f(J, t).
\end{equation}

The spatial operator is the variable-order symmetric Riesz operator $\mathbb{R}^{q(J)}$. This operator, also referred to as the Riesz-Weyl operator $\nabla^q$ in the fractional calculus literature, is the correct generator for symmetric L\'{e}vy flights \citep[see Sec 3.5, Eq. 58 and Appendix A.2 of][]{MetzlerKlafter2000}. It is defined as a sum of left- and right-sided Riemann-Liouville (RL) derivatives,

\begin{equation} \label{eq:riesz_variable}
\mathbb{R}^{q(J)} f(J) = C_q(J) \left[ {}_{a}D_J^{q(J)} f(J) + {}_{J}D_b^{q(J)} f(J) \right],
\end{equation}

\noindent
where $C_q(J) = - [2 \cos(\pi q(J) / 2)]^{-1}$. For a domain $J \in [a, b]$, the left-sided RL derivative is

\begin{equation} \label{eq:RL_left}
{}_{a}D_J^{q(J)} f(J) = \frac{1}{\Gamma(2-q(J))} \frac{d^2}{dJ^2} \int_a^J (J-s)^{1-q(J)} f(s) ds.
\end{equation}

\noindent
The Riemann-Liouville (RL) derivatives are integro-differential operators, meaning their mathematical definition involves both an integral and a derivative, as shown in Eq.~(\ref{eq:RL_left}). This structure is fundamentally different from the standard Fokker-Planck operator $\LFP$ (Eq.~\ref{eq:FP_local}), which is a purely differential operator.

The fundamental difference lies in their \textit{locality}. The standard Fokker-Planck operator $\LFP$, as defined in Eq.~(\ref{eq:FP_local}), is a local, second-order differential operator. Its evaluation at a point $J$ in phase space depends exclusively on the properties of the distribution function $f$ within an infinitesimal neighborhood of $J$, specifically $f(J)$, $f'(J)$, and $f''(J)$. This structure mathematically describes a continuous process, or Markovian walk, which emerges from stochastic steps with finite variance, as governed by the Central Limit Theorem. In contrast, the Riemann-Liouville (RL) derivative, defined in Eq.~(\ref{eq:RL_left}), is a \textit{non-local}, integro-differential operator. Its evaluation at $J$ explicitly depends on the \textit{global} values of the distribution function $f(s)$ over the entire domain $s \in [a, J]$. This dependence is encoded in the integral $\int_a^J (J-s)^{1-q(J)} f(s) ds$, which functions as a power-law weighted sum over the function's history or spatial extent. This non-local structure is the mathematical consequence of the Generalized Central Limit Theorem, describing processes (L\'{e}vy flights) characterized by stochastic steps with infinite variance.

The RL derivative's integral (e.g., $\int_a^J ... f(s) ds$ in Eq.~\ref{eq:RL_left}) computes the change at $J$ by summing contributions from the distribution function $f(s)$ over the entire domain $s \in [a, J]$. This non-local structure is essential for modeling Resonant Relaxation (RR) as a L\'{e}vy flight. It mathematically represents the large, discrete ``jumps'' in angular momentum that characterize RR, allowing a star to move from a distant angular momentum $s$ to $J$ without passing through the intermediate phase space. As analyzed in Section~\ref{sec:schwarzschild}, this non-local property is what allows stars to bypass regions of locally suppressed diffusion, such as the hypothesised Schwarzschild Barrier.

The heterogeneity of the system refers to the fact that the physical properties of the stellar cusp are not uniform but vary with location (i.e., with angular momentum $J$). The model accounts for this by allowing the diffusion parameters to be state-dependent. The diffusion coefficient $D(J)$ quantifies the strength of the relaxation process, while the fractional order $q(J)$ quantifies the nature of the transport (i.e., the exponent of the power-law jump distribution, Eq.~\ref{eq:levy_tail}). A variable $q(J)$ allows the model to describe a system where the process may be strongly non-local (e.g., $q=1.5$) in one part of the phase space and transition toward local diffusion (where $q \to 2$) in another.

\section{Timescales and Steady-State Implications}

The ratio of the diffusion coefficient for RR ($\DRR$) to that of classical two-body relaxation ($\Drlx$) is
\begin{equation} \label{eq:ratio}
\frac{\DRR(r)}{\Drlx(r)} = \frac{M_{\text{BH}}}{M_{\text{cusp}}(r)}.
\end{equation}
\noindent
RR dominates within the radius of influence $r_{\text{inf}}$.

\noindent
We analyze the steady-state regime, where the distribution function $f(J)$ is time-invariant ($\partial f / \partial t = 0$). This regime applies to systems in long-term equilibrium or in regions of phase space far from boundaries or barriers, where transport is not limited by the non-local effects discussed in Sections \ref{sec:transient} and \ref{sec:schwarzschild}. In this limit, the fractional framework must recover the known results of classical Resonant Relaxation (RR) theory, which serves as a consistency check for the model.

The transport process of RR, modeled as a L\'{e}vy flight, is inherently scale-free due to the power-law nature of its jump distribution (Eq.~\ref{eq:levy_tail}). This contrasts with Gaussian diffusion, which possesses a characteristic scale (the standard deviation). This scale-free property of the L\'{e}vy flight naturally derives the fundamental condition of steady-state RR, which is a constant flux $\mathcal{F}$ of stars through angular momentum space, $\mathcal{F}(J) \approx C(a)$. The flux $C(a)$ depends on the semi-major axis $a$ but is constant with respect to $J$.

As an illustration, we consider the population of stars $N(<J)$ with angular momentum less than $J$. In a steady state, the rate at which stars diffuse into this region (across $J$) must balance the rate at which they are removed (e.g., at the loss cone, $J \to 0$). The L\'{e}vy flight's long-range jumps ensure this balance results in a constant $\mathcal{F}(J)$. This constant flux condition, in turn, dictates the equilibrium profile $f(J)$. When this distribution is expressed in terms of orbital parameters such as eccentricity $e$ (where $1-e \propto J^2$) or the loss cone angle $\beta$ (where $\beta \propto J$), the constant flux solution directly yields the standard power-law distributions for stars near the loss cone,
\begin{equation} \label{eq:emri_ecc_dist}
P(1-e) \propto (1-e)^{-1/2}, \quad P(\beta) \propto \beta^{-3/2}.
\end{equation}
\noindent
Thus, the fractional formalism provides a more fundamental derivation for these known distributions, rooting them in the L\'{e}vy dynamics.

When we apply this framework to a specific astrophysical model, such as a strongly mass-segregated stellar cusp (a $\gamma=2$ profile, $n(r) \propto r^{-2}$, typical for heavy stellar-mass black holes), we can derive the event rates. The total event rate $\GRR$ (e.g., EMRIs) is the integral of the flux over the contributing semi-major axes. We find that the differential event rate scales as $d\GRR/da \propto a^{-3/2}$. Furthermore, comparing the magnitude of this RR-driven flux to the flux from classical two-body relaxation ($\Grlx$) yields the enhancement factor $E = \GRR / \Grlx$. This enhancement, $E \approx \sqrt{r_{\text{inf}}/r_{\text{min}}}$, where $r_{\text{inf}}$ is the SMBH influence radius and $r_{\text{min}}$ is the inner cutoff radius, quantifies the dominance of RR over two-body relaxation (Eq.~\ref{eq:ratio}) and is correctly recovered by the fractional model in this steady-state limit.

\section{Limitations of the Local Formalism for RR}
\label{sec:inconsistency}

The established approach in the literature involves modeling RR by calculating effective diffusion coefficients ($\DRR$) and inserting them into the classical, local Fokker-Planck equation,
\begin{equation} \label{eq:FP_local}
\frac{\partial f}{\partial t} = \LFP f = \frac{\partial}{\partial J} \left( \DRR(J) \frac{\partial f}{\partial J} \right).
\end{equation}
\noindent
This methodology includes sophisticated treatments such as the $\boldsymbol{\eta}$-formalism developed by \cite{BarOrAlexander2014} and utilized in \cite{BarOrAlexander2016}, which derives DCs by accounting for temporal correlations and the effects of precession (Adiabatic Invariance). Whilst these approaches provided foundational insights into the dynamics of RR under precession, we note that employing a local operator ($\LFP$) is not fully consistent with the underlying non-local physics of RR when modeled as a L\'{e}vy flight.

\subsection{The Nature of the Stochastic Process}
The limitation stems from the characterization of the stochastic process. Eq.~(\ref{eq:FP_local}) fundamentally describes Brownian motion or generalized Brownian motion (if noise is correlated but Gaussian), which relies on the Central Limit Theorem (CLT) and requires stochastic steps with finite variance. The physical description of RR adopted here corresponds to a L\'{e}vy flight (Eq.~\ref{eq:levy_tail}), a process where the variance of the step size is infinite {(Section~\ref{sec:justification})}. The evolution of such a system converges to a L\'{e}vy stable distribution (Generalized Central Limit Theorem, GCLT), rather than a Gaussian distribution.

While treatments such as \citet{BarOrAlexander2014} correctly handle the temporal correlations of the noise, they explicitly assume the noise amplitude is Gaussian. In their formalism \citep[Section 2.2 of][]{BarOrAlexander2014}, the background potential terms ($h_{nm}^{l}$) are described as time-dependent Gaussian random variables by invoking the CLT, justified by the superposition of forces from $N\gg1$ stars. {As argued in Section~\ref{sec:justification}, this justification is potentially flawed because the CLT requires finite variance of the individual contributions. If the variance of gravitational torques is infinite, the GCLT applies, leading to L\'{e}vy statistics, regardless of how large $N$ is.} This assumption inherently maintains finite variance for the resulting stochastic process and necessitates a description via a local FP equation. This contrasts with the L\'{e}vy flight model where the GCLT applies.

Furthermore, one may argue that a pure power-law jump distribution (Eq.~\ref{eq:levy_tail}) is unphysical, as it allows for arbitrarily large jumps. A more physical model might involve ``tempered'' L\'{e}vy flights, where the power law is suppressed by an exponential factor, $P(\Delta J) \sim |\Delta J|^{-1-q} e^{-\lambda J}$ \citep{Meerschaert2011StochasticMF}. This process, which converges to a ``tempered stable'' distribution, possesses finite variance. However, it is still fundamentally non-local and is described by a tempered space-fractional equation utilizing a tempered fractional operator $\mathbb{D}_x^{\alpha,\lambda}$ \citep{Meerschaert2011StochasticMF}. Crucially, this operator remains an integro-differential operator, maintaining the non-local structure required to model the jumps. The local Fokker-Planck equation, which imposes a Gaussian propagator, is therefore not the correct description even for this more physically realistic, non-local process. This reinforces our central claim: a local differential operator is {likely an inappropriate choice} for describing L\'{e}vy-like dynamics, whether pure or tempered.

\subsection{Operators: Locality vs. Non-locality}
The standard operator $\LFP$ is a local, second-order differential operator. This mathematical structure inherently requires particles to diffuse sequentially through adjacent points in phase space, prohibiting long-range ``jumps''. The appropriate operator for a L\'{e}vy flight, the fractional operator $\LFrac$, is an integro-differential operator (Section~\ref{subsec:superdiffusion_formalism}). It possesses a fundamentally non-local structure that accounts for these jumps.

\subsection{Propagators: Gaussian vs. L\'{e}vy Stable}
The fundamental solutions (propagators) highlight the limitations of the local model. We analyze the propagators in Fourier space, assuming constant coefficients $D$ and $q$. The Fourier transform (symbol) of the local operator $\LFP = D \partial^2/\partial J^2$ is $\hat{\mathcal{L}}_{\text{FP}}(k) = -D k^2$. The Fourier transform of the symmetric Riesz operator $\LFrac$ is $\hat{\mathbb{D}}^q(k) = -D |k|^q$.

\begin{figure*}
    \centering
    \includegraphics[width=0.98\textwidth]{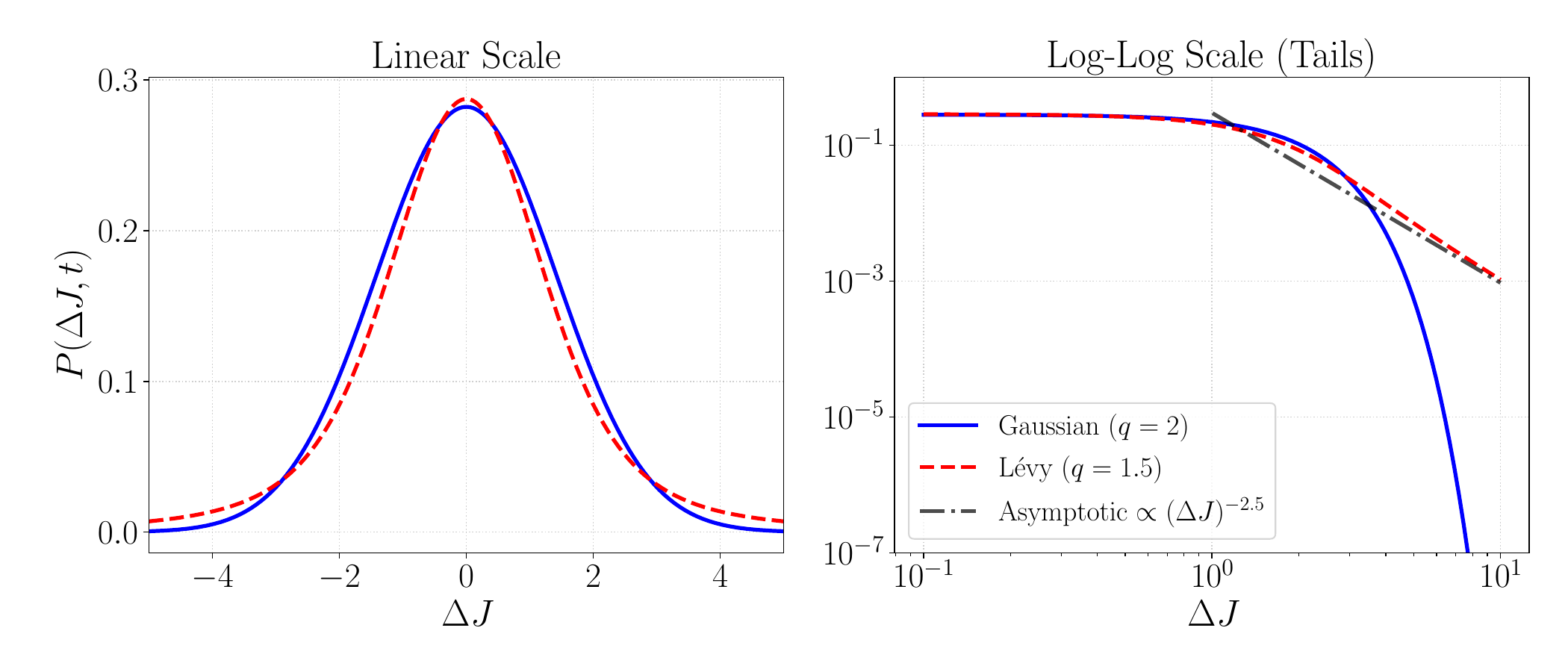}
    \caption{Comparison of the fundamental solutions (propagators) for local diffusion (Gaussian, $q=2$) and superdiffusion (L\'{e}vy stable, $q=1.5$), assuming $D=1, t=1$. Left panel (linear scale) shows the overall shape. Right panel (log-log scale) highlights the behavior of the tails. The Gaussian propagator exhibits rapid exponential decay (appearing parabolic in log-log). The L\'{e}vy propagator possesses a heavy power-law tail (linear shape in log-log, following the asymptotic behavior $(\Delta J)^{-2.5}$). This heavy tail significantly enhances the probability of non-local jumps (L\'{e}vy flights); for instance, at $|\Delta J|=5$ ($\approx 3.5\sigma$), the L\'{e}vy probability density is approximately 13 times higher than the Gaussian probability density.}
    \label{fig:propagators}
\end{figure*}

The propagator in Fourier space is $\hat{P}(k, t) = \exp(\hat{\mathcal{L}} t)$. The local model yields
\begin{equation} \label{eq:propagator_fourier_local}
\hat{P}_{\text{Gauss}}(k, t) = \exp(-D k^2 t).
\end{equation}
\noindent
In real space, this is the Gaussian propagator,
\begin{equation} \label{eq:propagator_gaussian}
P_{\text{Gauss}}(\Delta J, t) = \frac{1}{\sqrt{4\pi D t}} \exp\left(-\frac{(\Delta J)^2}{4Dt}\right).
\end{equation}
\noindent
The Gaussian propagator has exponentially suppressed tails (Figure~\ref{fig:propagators}).

The fractional model yields
\begin{equation} \label{eq:propagator_fourier_frac}
\hat{P}_{\text{Lévy}}(k, t) = \exp(-D |k|^q t).
\end{equation}
\noindent
In real space, this propagator $P_{\text{Lévy}}(\Delta J, t)$ has a power-law asymptotic behavior for large jumps,
\begin{equation} \label{eq:levy_tail_asymptotic}
P_{\text{Lévy}}(\Delta J, t) \sim D t (\Delta J)^{-1-q}.
\end{equation}
\noindent
This heavy tail mathematically captures the non-local L\'{e}vy flight (Figure~\ref{fig:propagators}, right panel). The contrast between the two propagators is stark. For the illustrative parameters used in Figure~\ref{fig:propagators} ($D=1, t=1, q=1.5$), the Gaussian standard deviation is $\sigma=\sqrt{2}$. The probability density of a large jump, e.g., $|\Delta J| = 5$ ($\approx 3.5\sigma$), is approximately 13 times higher for the L\'{e}vy propagator ($P_{\text{Lévy}} \approx 7.1 \times 10^{-3}$) than for the Gaussian propagator ($P_{\text{Gauss}} \approx 5.4 \times 10^{-4}$).

By adopting Eq.~(\ref{eq:FP_local}), the standard approach effectively imposes a Gaussian propagator onto a physical process governed by a L\'{e}vy propagator. This substitution inherently eliminates the non-local transport that characterizes RR, leading to divergent predictions in specific regimes, as detailed below.

\section{The Transient Regime: TDEs in Post-Starburst Galaxies}
\label{sec:transient}

The divergence between the predictions of local and fractional formalisms becomes particularly pronounced in transient, non-equilibrium scenarios. A crucial astrophysical application concerns the unexpectedly high rate of Tidal Disruption Events (TDEs) observed in ``E+A'' or post-starburst galaxies. These galaxies exhibit spectral signatures indicating a major starburst event that concluded within the last $\sim 1$ Gyr, often associated with galaxy mergers \citep[see e.g.][]{ArcaviEtAl2014,FrenchEtAl2016,LawSmithEtAl2017}.

Such major perturbations likely disrupt the stellar cusp and empty the loss cone around the central SMBH. The observed high TDE rates in these systems therefore imply that the loss cone must be repopulated rapidly following the perturbation. This scenario presents a transient problem where the specific nature of the angular momentum transport mechanism---local diffusion versus non-local L\'{e}vy flights---dictates the timescale of the response.

\subsection{Transient Loss Cone Filling: Exponential Delay vs. Immediate Response}
We analyze the refilling of an initially empty loss cone. We examine the population near the loss cone ($J \to 0$) due to transport from the bulk of the distribution at $J' \approx J_{\text{orb}}$. We define the characteristic diffusion timescale as $\Tdiff$.

\begin{figure}
    \centering
    \includegraphics[width=\columnwidth]{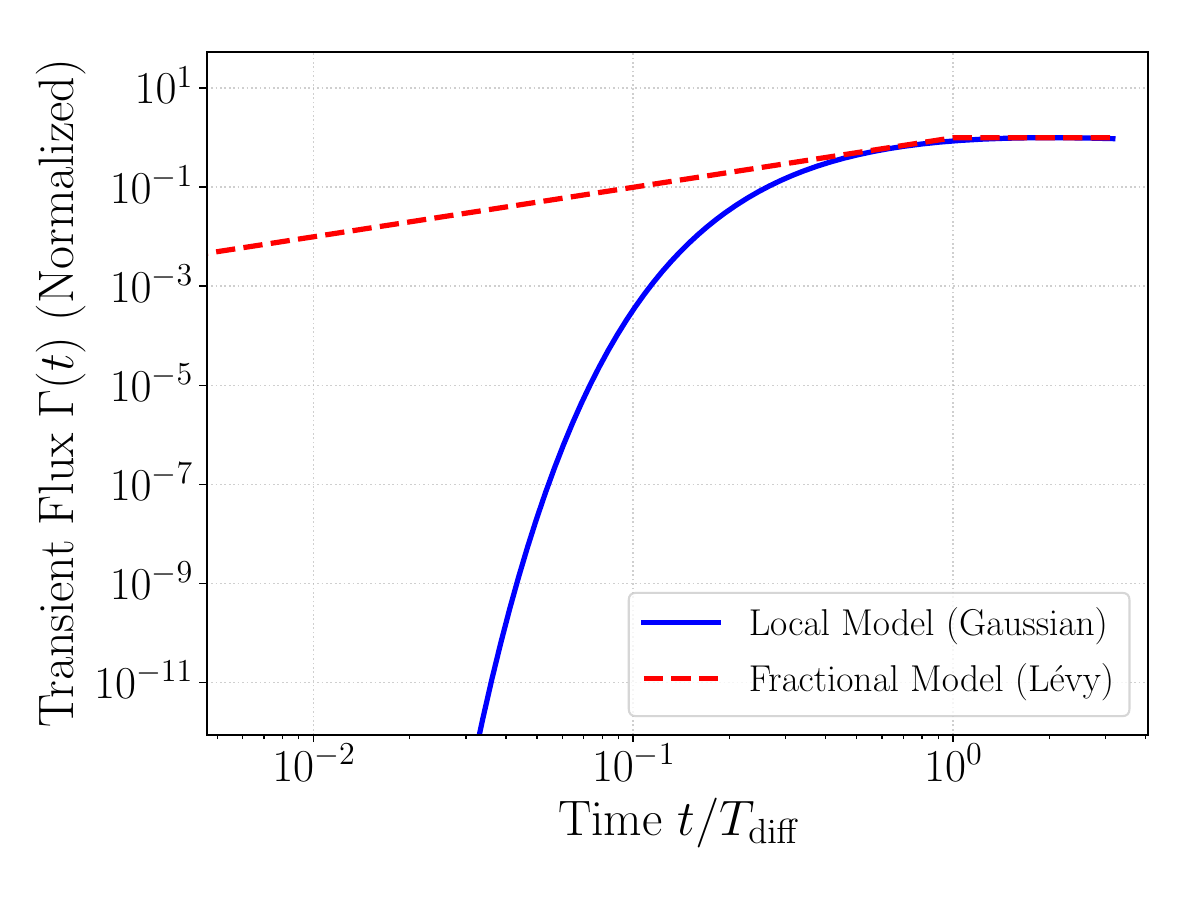}
    \caption{Evolution of the transient flux into the loss cone following a perturbation, normalized by the diffusion timescale $\Tdiff$. The log-log scale highlights the behavior at early times. The local model (blue line) exhibits exponential suppression ($\Gamma(t) \propto \exp(-C/t)$), leading to a significant delay. It reaches 1\% of the steady-state flux only at $t \approx 0.17\, \Tdiff$. The fractional model (red dashed line) exhibits immediate linear growth ($\Gamma(t) \propto t$) due to non-local jumps, reaching the 1\% flux level approximately 17 times faster at $t = 0.01\, \Tdiff$.}
    \label{fig:transient}
\end{figure}

In the classical (local) diffusion model, such as those based on the standard FP equation, the evolution is governed by the tail of the Gaussian propagator (Eq.~\ref{eq:propagator_gaussian}). The population near $J=0$ at time $t$ is
\begin{equation} \label{eq:transient_gauss}
f_{\text{Gauss}}(0, t) \sim t^{-1/2} \exp\left(-\frac{C}{t}\right).
\end{equation}
\noindent
For early times ($t \ll \Tdiff$), the population is exponentially suppressed. The local model mandates that stars diffuse step-by-step across the angular momentum space. It predicts a significant delay before the TDE rate rises (Figure~\ref{fig:transient}). In the simulation shown, at $t/\Tdiff = 0.01$, the normalized local flux is effectively zero ($\sim 10^{-43}$). The time required to reach 1\% of the steady-state flux is $t \approx 0.17\, \Tdiff$. This delay appears inconsistent with the high observed rates in post-starburst environments.

In the fractional (non-local) superdiffusion model, the evolution is governed by the power-law tail of the L\'{e}vy propagator (Eq.~\ref{eq:levy_tail_asymptotic}). This asymptotic form, $P_{\text{Lévy}}(\Delta J, t) \sim D t (\Delta J)^{-1-q}$, is the known propagator for the space-FFPE, corresponding to the asymptotic behavior $W(x,t) \sim K^\mu t |x|^{-1-\mu}$ for L\'{e}vy stable laws \citep[see Sec 3.5, Eq. 62 of][]{MetzlerKlafter2000}. The population near $J=0$ is therefore

\begin{equation} \label{eq:transient_levy}
f_{\text{Lévy}}(0, t) \sim D t (J_{\text{orb}})^{-1-q}.
\end{equation}

\noindent
The population increases linearly with time, starting immediately from $t=0$ (Figure~\ref{fig:transient}). The non-local nature of RR allows stars to ``jump'' directly from $J_{\text{orb}}$ to $J=0$ in a single dynamical timescale. In this model, the 1\% flux level is reached rapidly at $t = 0.01\, \Tdiff$, approximately 17 times faster than the local model.

The ratio of the transient fluxes at early times diverges dramatically,
\begin{equation} \label{eq:transient_ratio}
\frac{\Gamma_{\text{Lévy}}(t)}{\Gamma_{\text{Gauss}}(t)} \sim t^{3/2} \exp\left(+\frac{C}{t}\right).
\end{equation}
\noindent
The fractional framework provides the necessary physical mechanism (non-local L\'{e}vy flights) to explain the rapid replenishment of the loss cone and the high TDE rates observed in these non-equilibrium galaxies.

\section{The Nullification of the proposed Schwarzschild Barrier}
\label{sec:schwarzschild}

A second critical area where the local formalism yields substantially different results concerns the impact of relativistic precession on EMRI rates. This effect has been conceptualized and put forward as the Schwarzschild Barrier (SB), which, in the standard view \textit{based on local models}, represents a major bottleneck in EMRI formation theory.

\subsection{The Schwarzschild Barrier Hypothesis in the Adiabatic Invariance and Local Suppression}

RR relies on the coherence of torques. As a star approaches the SMBH, General Relativistic (GR) precession becomes rapid, decoupling the star from the coherent torques and quenching RR. This defines a critical angular momentum $\JSB$. For $J < \JSB$, the efficiency of RR is locally suppressed.

Classical two-body relaxation (TBR) is driven by stochastic encounters and does not depend on orbital coherence. It is entirely unaffected by the SB. However, TBR is much slower than RR in realistic nuclei.

The mechanism for RR suppression is understood as Adiabatic Invariance (AI). As detailed by \cite{BarOrAlexander2014}, when the precession frequency exceeds the maximal variability frequency of the (assumed smooth and Gaussian) background noise, the effective diffusion coefficient is exponentially suppressed. This suppression was observed in N-body simulations using Post-Newtonian approximations \citep{MerrittEtAl11} {and \citet{HamersEtAl2014},} and interpreted as a barrier. The standard view, based on these local models, is that the SB creates a significant bottleneck, as transport across the barrier region is limited primarily to the slow rate of two-body relaxation \citep{BarOrAlexander2016}.

\subsection{Limitations of Local Models at the Barrier}
The standard local model (Eq.~\ref{eq:FP_local}), including the implementation of the $\boldsymbol{\eta}$-formalism by \cite{BarOrAlexander2016}, implements the SB by sharply reducing the diffusion coefficient $D(J)$ for $J < \JSB$ (see Figure~\ref{fig:barrier}). In a local formalism, the steady-state flux is determined by the integrated resistance, $\mathcal{F}_{\text{local}} \propto (\int dJ/D(J))^{-1}$. Particles must pass sequentially through all intermediate points.

Consequently, the local model necessarily predicts that the flux is severely reduced. In the scenario depicted in Figure~\ref{fig:barrier}, the diffusion coefficient inside the barrier ($J < \JSB$) is suppressed by 95\% (reduced to the two-body relaxation floor of 0.05, a factor of 20 reduction). This high resistance dominates the integral, reducing the steady-state local flux globally to only $\approx 12\%$ of the flux achievable if the barrier were absent (a reduction factor of $\approx 8.3$).

\subsection{Fractional Dynamics: Non-Local Barrier Jumping}
The fractional framework provides a fundamentally different result due to the non-local nature of the transport operators. This distinction {potentially resolves} the bottleneck suggested by the local models. We demonstrate this mathematically by analyzing the structure of the fractional derivatives.

We consider the right-sided Riemann-Liouville (RL) derivative (defined in Section~\ref{subsec:superdiffusion_formalism}), which governs transport from higher $J$ to lower $J$. For a domain $[a, b]$, and assuming constant $q$, the operator structure (similar to Eq.~\ref{eq:RL_left}) is:
\begin{equation} \label{eq:RL_right}
{}_{J}D_b^{q} f(J) \propto \frac{d^2}{dJ^2} \int_J^b (s-J)^{1-q} f(s) ds.
\end{equation}
\noindent
Let us evaluate this operator at a point $J_{\text{in}} < \JSB$ (inside the barrier region). We split the integral at the barrier boundary $\JSB$,
\begin{align} \label{eq:Integral_split}
I(J_{\text{in}}) &= \int_{J_{\text{in}}}^{\JSB} (s-J_{\text{in}})^{1-q} f(s) ds + \nonumber \\
& \quad \int_{\JSB}^{b} (s-J_{\text{in}})^{1-q} f(s) ds.
\end{align}
\noindent
The first term represents transport originating locally within the barrier region, where diffusion is suppressed.

The second term represents non-local transport originating from outside the barrier ($s > \JSB$), where RR is strong. This term mathematically models L\'{e}vy flights that start outside the barrier and ``jump over'' the quenched region $J < \JSB$, landing directly at $J_{\text{in}}$.

This ``barrier jumping'' mechanism is inherent to the integral structure of the fractional operator. It is mathematically excluded by the differential structure of the local Fokker-Planck formalism.

\subsection{The Nullification of the Barrier and Reinterpretation of Prior Results}
\begin{figure}
    \centering
    \includegraphics[width=\columnwidth]{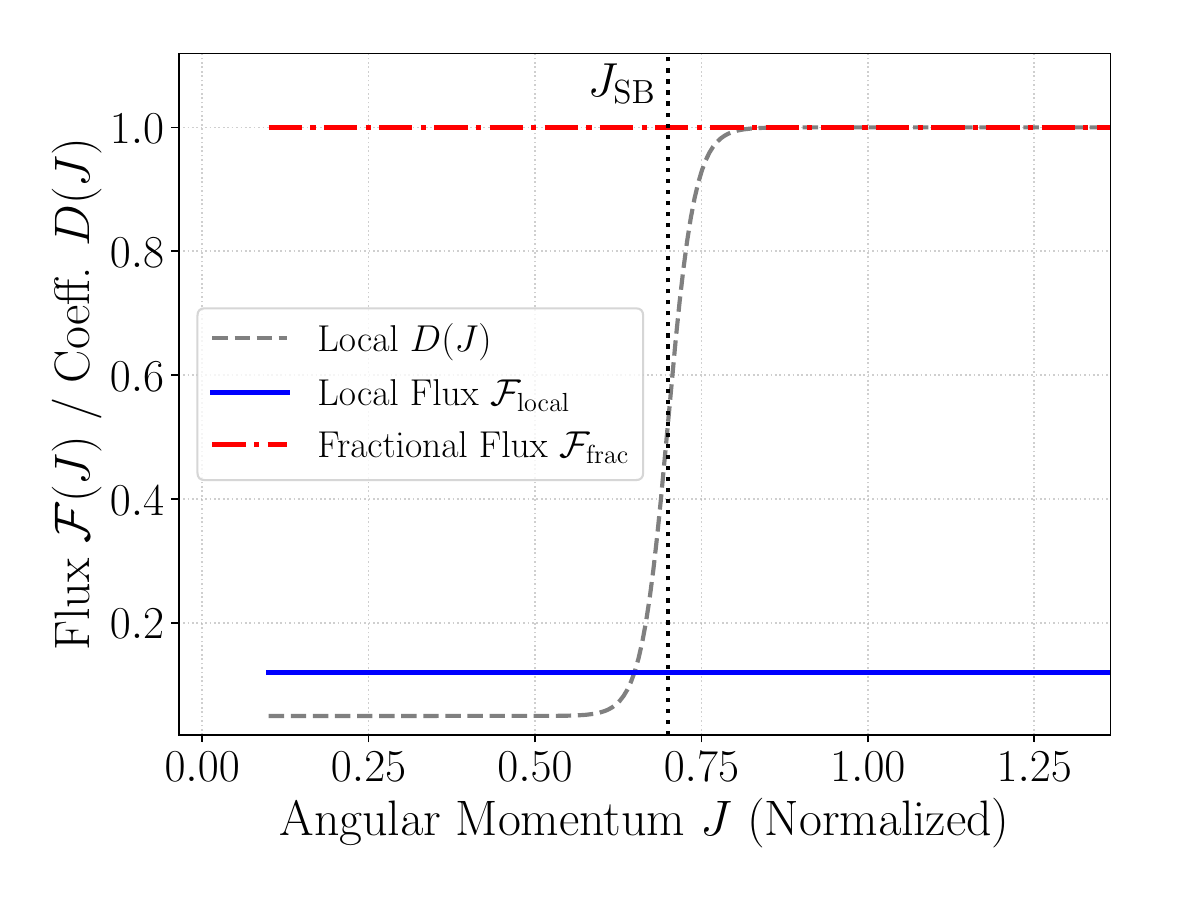}
    \caption{Impact of the Schwarzschild Barrier ($\JSB$) on the steady-state angular momentum flux $\mathcal{F}(J)$. The local diffusion coefficient $D(J)$ (gray dashed line) drops by 95\% (a factor of 20) for $J < \JSB$ due to Adiabatic Invariance. In the local model (blue solid line), the flux is limited by the integrated resistance ($\int 1/D(J) dJ$), resulting in an 88\% reduction of the global flux ($\mathcal{F}_{\rm local} \approx 0.12$). The fractional model (red dash-dotted line) shows that the flux is maintained ($\mathcal{F}_{\rm frac} = 1.0$) by non-local ``barrier jumping'' from $J > \JSB$. The fractional flux is $\sim 8.3$ times higher than the local flux.}
    \label{fig:barrier}
\end{figure}

The consequence of this non-local transport is profound. The bottleneck predicted by local models does not inhibit the flux in a non-local framework. The flux into the loss cone is maintained by direct jumps from the region $J > \JSB$. As shown in Figure~\ref{fig:barrier}, the fractional flux remains unsuppressed ($\mathcal{F}_{\text{frac}}=1.0$), completely overcoming the 88\% reduction predicted by the local model. Therefore, the proposed Schwarzschild Barrier, while representing a real physical phenomenon of local quenching via AI, does not constitute an actual barrier to RR-driven transport {in this framework}.

{The apparent discrepancy between this theoretical result and the numerical observations of a barrier in \citet{MerrittEtAl11} (MAMW11) and \citet{HamersEtAl2014} warrants careful consideration. We argue this discrepancy stems primarily from the methodologies used to analyze the simulations, which inherently assume local transport.}

{\citet{HamersEtAl2014} analyzed the dynamics by measuring the diffusion coefficients (DCs), $D(L) = \langle (\Delta L)^2 \rangle / \Delta t$. They correctly observed a sharp suppression (a "knee") in $D(L)$ near the SB. However, if RR is a L\'{e}vy flight, the second moment $\langle (\Delta L)^2 \rangle$ formally diverges (Section~\ref{sec:justification}). In any finite simulation, the measured moments are necessarily finite because the power-law tail is truncated. The measured $D(L)$ is therefore an \textit{effective} diffusion coefficient, dominated by the frequent small steps and under-representing the rare large jumps \citep{MetzlerKlafter2000}.}

{The suppression measured by \citet{HamersEtAl2014} reflects the quenching of \textit{local} transport (small steps) due to Adiabatic Invariance. The crucial non-local jumps, which bypass the barrier, are too infrequent to significantly contribute to their measurement of $D(L)$ over the analyzed timescales. Consequently, analyzing the system via local diffusion coefficients leads to the interpretation that the overall transport is suppressed, even when non-local flux persists.}

{In the case of MAMW11, the barrier was identified visually by the "bouncing" of trajectories. This is expected even in the fractional framework, as most steps are small and affected by local quenching. Furthermore, MAMW11 utilized very small N ($N=50$), enhancing TBR relative to RR. MAMW11 concluded that the few captures they observed were likely driven by TBR, not RR.}

Since RR can bypass the quenched region via non-local jumps, and two-body relaxation is inherently unaffected by precession, we conclude that the Schwarzschild Barrier {may not impose the significant inhibition previously assumed} on the production of EMRIs \citep{Amaro-SeoaneLRR,Amaro-Seoane2020,Amaro-SeoaneEtAl07}. This result {addresses} a major theoretical bottleneck in EMRI formation theory.

\section{Numerical Evidence from N-body Simulations}
\label{sec:simulations}

To {test} the theoretical prediction that the Schwarzschild Barrier (SB) is ineffective against non-local transport, we analyze the results of direct-summation $N$-body simulations incorporating relativistic corrections, following the pioneering work of \citep{KupiEtAl06}. The full details of the numerical methods, including the post-Newtonian (PN) terms, are available in \citet{BremEtAl2014}.

\subsection{Simulation Setup and Methodology}

The simulations model a central SMBH with a mass of $M_\bullet = 10^6 \msun$ surrounded by $N=400$ stellar-mass black holes (BHs), each with a mass of $m_\star = 50 \msun$. The stellar BHs are initialized with a flat distribution in semi-major axis $a$ and $e^2$ {(a thermal distribution)}, with $a$ ranging from $0.1 \rm{mpc}$ to $10 \rm{mpc}$.

{These simulations are intended as a proof of concept and possess significant limitations. The particle number ($N=400$) is very small compared to realistic galactic nuclei, and the individual stellar mass ($m_\star = 50 \msun$) is large. This significantly enhances the efficiency of two-body relaxation (TBR) relative to RR. Furthermore, the initialization with a thermal eccentricity distribution means the system starts near equilibrium, rather than demonstrating the dynamic process of diffusing across the barrier from low eccentricities.}

This range of semi-major axes is crucial. For a $10^6 \msun$ SMBH, the radius of influence $r_{\rm inf}$ is typically on the order of parsecs. The simulated domain (milliparsecs) is therefore deep within the SMBH's sphere of influence. In this region, the enclosed stellar mass $M_{\text{cusp}}(r)$ is much smaller than $M_{\bullet}$. According to Eq.~(\ref{eq:ratio}), this ensures that Resonant Relaxation (RR) is the dominant transport mechanism, significantly outweighing two-body relaxation{, provided N is sufficiently large and $m_\star$ sufficiently small}. Simultaneously, this proximity is where relativistic precession becomes rapid, defining the location of the SB.

The figures presented here are generated from a statistical sample of $N_{\rm simul}=300$ such simulations. {The simulations are run for a duration sufficient to establish a quasi-steady state distribution.} We generate a 2D histogram by sampling the orbital elements (semi-major axis $a$ and eccentricity $e$) of all $N=400$ particles across all 300 simulations throughout their evolution. The orbital elements are binned into a $100 \times 100$ grid in $\log_{10}(a)$ versus $\log_{10}(1-e)$ phase space. The count in each bin is then normalized by the total number of samples and the bin area to create a presence density map, $\log_{10} P(a, 1-e)$, which represents the time-averaged probability of finding a particle in a given region of phase space.

\subsection{Results and Interpretation}

\begin{figure*}
    \centering
    \includegraphics[width=0.45\textwidth]{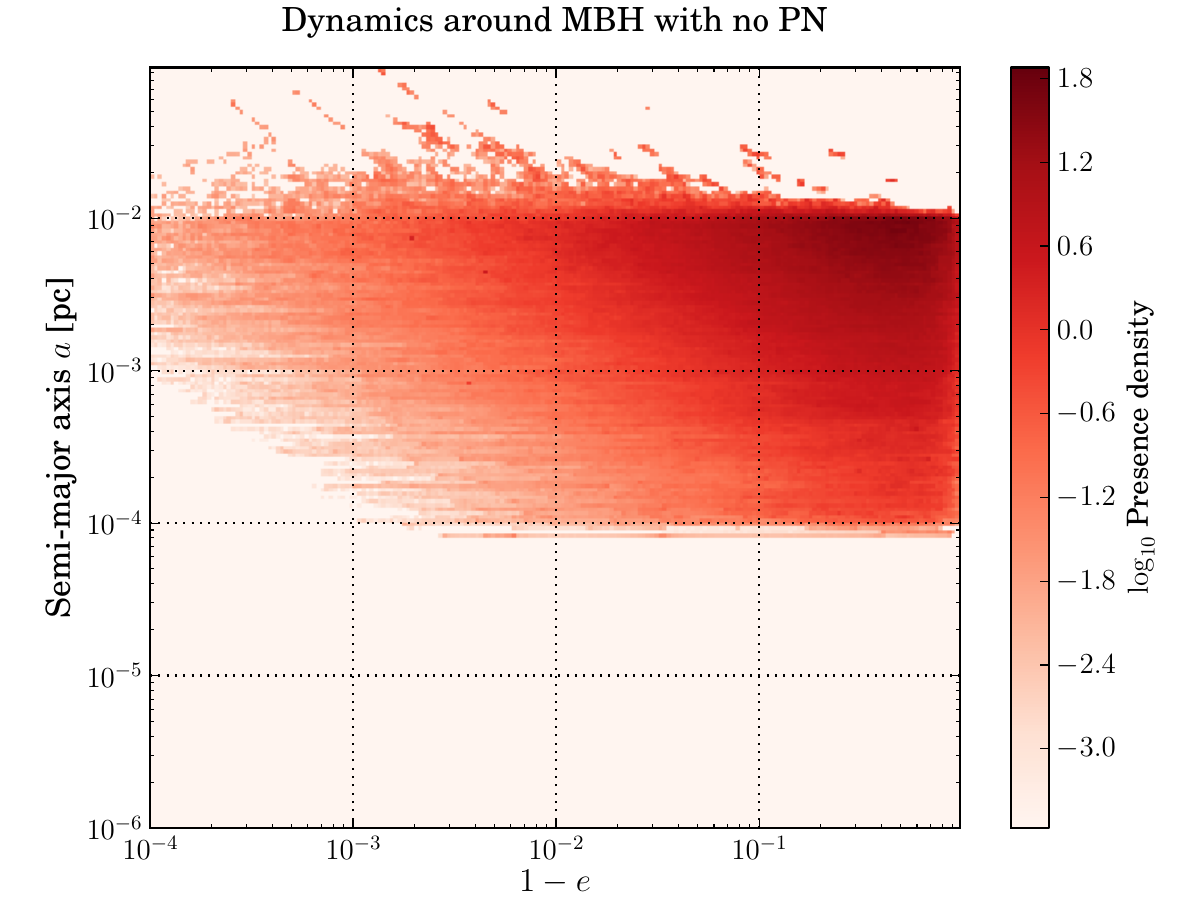}
    \includegraphics[width=0.45\textwidth]{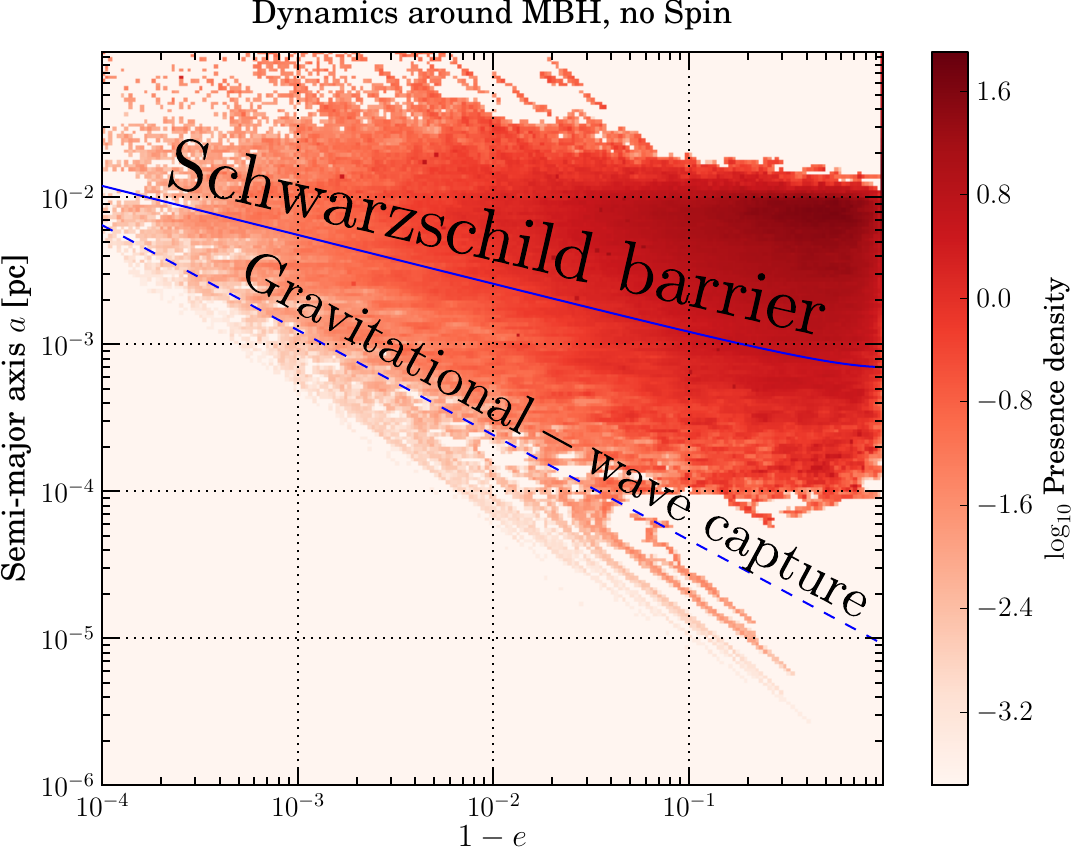}
    \caption{Time-averaged presence density of stellar-mass black holes in the $a$ vs. $1-e$ phase space, generated from 300 $N$-body simulations of 400 particles each, focusing on the critical milliparsec scales. \textit{Left panel:} The purely Newtonian case (no PN terms). Particles are driven to high eccentricities (low $1-e$) by relaxation processes, filling the phase space. \textit{Right panel:} The relativistic (PN) case without spin. The solid blue line marks the theoretical location of the Schwarzschild Barrier ($\tau_{\rm SS} \approx \tau_{\rm coh}$), and the dashed blue line marks the gravitational wave capture limit ($\tau_{\rm GW} < \tau_{\rm rlx}$). The particle density is high on \textit{both} sides of the SB (solid line), with no visible depletion or barrier to transport in this TBR-dominated regime.}
    \label{fig:histo_comparison}
\end{figure*}

In Figure~\ref{fig:histo_comparison}, we present the resulting presence density maps. The left panel shows the purely Newtonian simulation (no PN terms), which serves as a baseline. In the absence of relativistic precession, RR operates efficiently, driving particles toward high eccentricities (low $1-e$) and densely populating the phase space.

The right panel shows the full post-Newtonian case (without spin). In this plot, the solid blue line indicates the theoretical location of the Schwarzschild Barrier, where the Schwarzschild precession timescale equals the coherent RR timescale ($\tau_{\rm SS} \approx \tau_{\rm coh}$). The dashed blue line indicates the gravitational wave (GW) capture limit, where the GW inspiral timescale becomes shorter than the two-body relaxation timescale ($\tau_{\rm GW} < \tau_{\rm rlx}$) \citep{Amaro-SeoaneLRR}.

It is visually evident from the right panel that the Schwarzschild Barrier does not function as an actual barrier to transport {in these simulations}. The presence density (dark red shading) is high on both sides of the solid blue line. 

{However, the interpretation of this result requires careful consideration of the relaxation regimes, as analyzed by \citet{HamersEtAl2014}. They demonstrated that the impact of the SB depends on the relative strength of TBR versus the suppressed RR (which they termed Anomalous Relaxation, AR) below the barrier. They defined a critical radius, $a_{\mathrm{AR,max}}$ above which TBR is strong enough to dominate transport across the SB. For the parameters used here ($N=400, m_\star=50\msun, M_\bullet=10^6\msun$), we estimate $a_{\mathrm{AR,max}} \approx 1.4$ mpc.}

{Therefore, for the majority of the simulated phase space ($a > 1.4$ mpc), the efficient transport observed across the SB is likely dominated by TBR, which is inherently unaffected by relativistic precession. The simulations are in a regime where enhanced TBR (due to low N and high $m_\star$) masks the effects of the SB on RR. Consequently, while these results show efficient transport, they cannot be used to conclusively validate the hypothesis that \textit{RR} bypasses the barrier via non-local jumps.}

The particle distribution is not truncated or depleted at the SB. Instead, the density extends all the way to the GW capture limit (dashed line), where particles are removed from the system. This provides {numerical evidence of efficient transport across the SB.}

{To robustly validate the fractional diffusion model, a comprehensive numerical investigation is required. This investigation must include simulations with significantly larger $N$ and smaller $m_\star$ to ensure the dominance of RR and reduce the impact of TBR (ensuring $a \ll a_{\mathrm{AR,max}}$), exploration of different initial conditions (e.g., starting from quasi-circular orbits far from the barrier), and, critically, a direct measurement of the angular momentum jump probability distribution $P(\Delta J)$ from the simulations. This measurement is essential to confirm the presence of the heavy tails predicted by the L\'{e}vy flight hypothesis (Section~\ref{sec:justification}). Such an extensive study requires significant computational resources and specialized analysis techniques; it is beyond the scope of this theoretical paper and will be presented in a separate, dedicated work.}

\section{Conclusions}

{We have investigated the application of a fractional dynamics framework for modeling transport in stellar nuclei. We provided a formal justification (Section~\ref{sec:justification}) that Resonant Relaxation (RR) must be characterized as a L\'{e}vy flight due to the infinite variance inherent in the gravitational torques.} {We utilized established results from statistical physics \citep[e.g.,][]{Meerschaert2011StochasticMF} that superdiffusive processes (L\'{e}vy flights) are characterized by infinite variance and heavy-tailed jump distributions.} This physical reality contravenes the assumptions of the standard local Fokker-Planck equation, which is mathematically restricted to Gaussian processes governed by the Central Limit Theorem. By applying the space-Fractional Fokker-Planck Equation (space-FFPE), which employs non-local integro-differential operators (Eq.~\ref{eq:riesz_variable}), we {provide a model consistent with} the Generalized Central Limit Theorem.

{This fractional framework offers potential resolutions to significant discrepancies between local theory and astrophysical observations.} We showed that local models predict {a significant} exponential delay in transient loss cone refilling ($\Gamma(t) \propto \exp(-C/t)$), failing to explain high TDE rates in post-starburst galaxies. The fractional model, by contrast, predicts an immediate linear flux ($\Gamma(t) \propto t$) (Section~\ref{sec:transient}).

Furthermore, we {demonstrated analytically} that the non-local integral structure of the fractional operators permits ``barrier jumping'' (Section~\ref{sec:schwarzschild}). This mechanism {bypasses} the postulated Schwarzschild Barrier as a transport bottleneck. {We presented preliminary, proof-of-concept direct $N$-body simulations} incorporating relativistic effects (Section~\ref{sec:simulations}), which {indicate} that the particle presence density is continuous across the theoretical barrier. {However, we analyzed the relaxation regimes following \citet{HamersEtAl2014} and concluded that this observed efficiency is likely due to enhanced TBR (due to small $N$ and large $m_\star$) rather than non-local RR. Further investigation with large-N simulations, including direct measurement of the jump statistics, is required to validate the non-local RR hypothesis and will be presented in a forthcoming work.}

Local models, bound by differential operators, misinterpret this local quenching of diffusion as a severe barrier to EMRI production \citep{BarOrAlexander2016}. Our theoretical findings {strongly suggest} that if RR is indeed a L\'{e}vy flight, this barrier may not inhibit EMRI rates {as severely as previously thought}, {potentially resolving} a major theoretical bottleneck.

The implementation of this fractional framework into existing stellar dynamics codes presents a significant but necessary numerical challenge. Standard Fokker-Planck solvers are typically built on finite-difference schemes, such as the Chang-Cooper method, which discretize a local differential operator ($\LFP$). This results in sparse (e.g., tridiagonal) matrices that are computationally efficient to solve. The fractional operators, however, are non-local integro-differential operators. Their discretization, for instance, can be achieved using a fractional finite difference method. This method approximates the Riesz operator $\mathbb{D}_x^\alpha f(x)$ by $h^{-\alpha} \Delta^\alpha f(x)$, where $\Delta^\alpha$ is a fractional centered-difference operator derived from the Grünwald-Letnikov series $\Delta^\alpha f(x) = \sum_{j=0}^\infty \binom{\alpha}{j} (-1)^j f(x-jh)$ \citep{Meerschaert2011StochasticMF}. This discretization, which couples $f(x)$ to all points $f(x-jh)$, is what leads to the dense matrix system and its $O(N^2)$ (or even worse) scaling. However, stable implicit finite difference codes can be constructed, for example by shifting the index of the numerical operator, which is essential for robust numerical solutions \citep{Meerschaert2011StochasticMF}. Despite this computational cost, this transition {appears warranted}. The results of this work demonstrate that local approximations {may be qualitatively incorrect, potentially predicting} spurious bottlenecks and time delays that are artifacts of the model rather than the physics. Adopting these non-local numerical methods is therefore imperative for producing physically {accurate} simulations of event rates in galactic nuclei.

\end{document}